\documentclass{article}
\pdfoutput=1
\usepackage[lmargin=0.75in,rmargin=0.75in,tmargin=1.0in,bmargin=1.0in]{geometry}
\usepackage{wrapfig}
\usepackage{amsmath}
\usepackage{wasysym}
\usepackage[affil-it]{authblk} 
\usepackage{graphicx}
\usepackage[font={small}]{caption} 

\usepackage[square,numbers,sort]{natbib}
\usepackage{authblk}

\usepackage{pdfpages}
\selectfont

\begin{document}

\title{\Large Isotopic profiles imply strong convective influence on water near the tropical tropopause}
\author[a]{Maximilien Bolot} 
\author[b]{Bernard Legras} 
\author[c]{Kaley A. Walker}
\author[d]{Christopher D. Boone}
\author[e]{Peter Bernath}
\author[f]{William G. Read} 
\author[a]{Elisabeth J. Moyer\footnote{E.J. Moyer. E-mail: moyer{@}uchicago.edu}}

\affil[a]{Dept.\ of the Geophysical Sciences, University of Chicago, Chicago, IL, USA}
\affil[b]{Laboratoire de M\'et\'eorologie Dynamique, CNRS and Ecole Normale Sup\'erieure, Paris, France}
\affil[c]{Dept.\ of Physics, University of Toronto, Toronto, ON, Canada}
\affil[d]{Dept.\ of Chemistry, University of Waterloo, Waterloo, ON, Canada}
\affil[e]{Dept.\ of Chemistry and Biochemistry, Old Dominion University, Norfolk, VA, USA}
\affil[f]{Jet Propulsion Laboratory, Pasadena, CA, USA}

\date{}

\maketitle

%
%
%

\abstract{\bf The influence of deep convection on water vapor in the Tropical Tropopause Layer (TTL), the region just below the high ($\sim$18 km), cold tropical tropopause, remains an outstanding question in atmospheric science. Moisture transport to this region is important for climate projections because it drives the formation of local cirrus (ice) clouds, which have a disproportionate impact on the Earth's radiative balance.
Deep cumulus towers carrying large volumes of ice are known to reach the TTL, but their importance to the water budget has been debated for several decades.
We show here that profiles of the isotopic composition of water vapor can provide a quantitative estimate of the convective contribution to TTL moistening. Isotopic measurements from the ACE satellite instrument, in conjunction with ice loads inferred from CALIOP satellite measurements and simple mass-balance modeling, suggest that convection is the dominant source of water vapor in the TTL up to near-tropopause altitudes.
The relatively large ice loads inferred from CALIOP satellite measurements can be produced only with significant water sources, and isotopic profiles imply that these sources are predominantly convective ice. 
Sublimating ice from deep convection appears to increase TTL cirrus by a factor of several over that expected if cirrus production were driven only by large-scale uplift; sensitivity analysis implies that these conclusions are robust for most physically reasonable assumptions. 
Changes in tropical deep convection in future warmer conditions may thus provide an important climate feedback.}



The tropical tropopause layer (TTL) is a few-kilometer region below the high tropical tropopause ($\sim$17--18 km) characterized by slow ascent of air, ultimately into the stratosphere \cite{Fueglistaler-2009}. The TTL is by definition a region where convective influence falls off sharply: the base of the TTL is commonly taken as  
the level of clear sky zero radiative heating, which divides the region of mean large-scale descent, where convective transport dominates, from that of mean ascent above. 
However, the contribution of convection to the TTL water budget has been debated for decades (e.g.\ \cite{Danielsen-1993, Jensen-2007, James-2009, Randel-2013}).

The TTL is also the location where air ascending to the stratosphere experiences its final drying, 
producing a blanket of thin high-altitude cirrus \cite{McFarquhar-2000} with a disproportionately large radiative impact, $\sim$20 W/m$^2$ locally and $\sim$4 W/m$^2$ in tropics-wide average 
\cite{Haladay-2009}.
The final water content of air entering the stratosphere appears set by the cold point tropopause temperature \cite{Bonazzola-2004,Fueglistaler-2005,James-2009},
but the volume of cirrus that dehydrates that air is governed by the total water transport to the TTL, since all added moisture must eventually be removed by ice formation and sedimentation \cite{Holton-2001}.

Overshooting convection is likely pervasive in the TTL: a recent study \cite{Tissier-2016} suggests that most air parcels at the tropical tropopause have been in recent contact with a cloud well above the mean level of neutral buoyancy. Trajectory-based studies typically assume that air parcels encountering these events gain water vapor to saturation \cite{Dessler-2007, James-2008,Tzella-2011,Bergman-2012}, and  
observations and simulations also suggest that convection does indeed moisten
the TTL \cite[e.g.][]{Jensen-2007,Dauhut-2014}.
However, the profile of convective water transport has been difficult to determine. Satellite images provide only rough guidance for cloud-top altitudes and no measure of how much detrained ice ultimately sublimates. 
Field campaigns over restricted areas (e.g.\ \cite{Corti-2008}) provide local information but
cannot be scaled up to the whole tropics.
Remote-sensing water vapor measurements cannot diagnose how much water is removed by cirrus formation or distinguish the different pathways by which water reaches the TTL.
Disentangling and quantifying the sources of TTL moisture requires widespread observations of some additional tracer. 

\begin{figure}[ht!!!]
\centering
\includegraphics[width=0.45\textwidth]{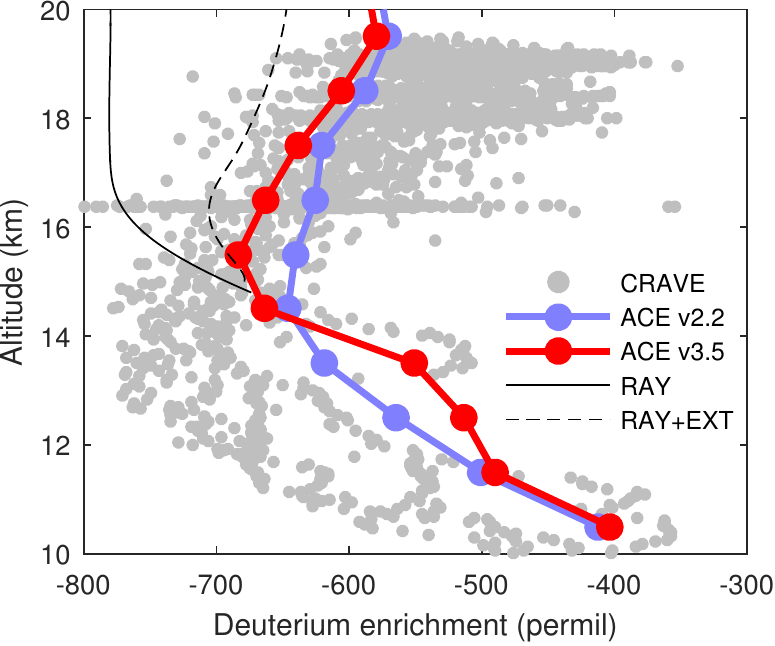}
\caption{Mean vertical profile of tropical (15$^\circ$S--15$^\circ$N) water vapor isotopic composition from ACE. We show both version 3.5 (red, \cite{Boone-2013}) and the prior version 2.2 (blue \cite{Boone-2005}); we use primarily v3.5 but show results from inverting both profiles. 
For comparison we also show (grey) in-situ measurements from the Harvard-ICOS instrument \cite{Sayres-2009} during the winter 2007 CR-AVE mission near Costa Rica (data restricted to 6$^\circ$N--11$^\circ$N). 
All measurements exhibit a similar general pattern, with a turnover of the isotopic profile at $\sim$14--15 km. This enhancement implies some source of heavy water to the TTL. 
Lines show predicted isotopic compositions in the absence of convective sources, for simple uplift (solid) and with extratropical mixing (dotted), when the prediction is started from the level of clear sky zero radiative heating. Neither case can reproduce observed profiles. 
\label{fig:fig1}}
\end{figure}

The isotopic composition of water (the ratio of heavy to light isotopologues, e.g.\ HDO/H$_2$O) is a relatively new tool that can serve this purpose.
Because the heavier isotopologues preferentially condense during ice formation \cite{Merlivat-1967,Jouzel-1984}, different moisture sources or removal processes are associated with different isotopic signatures. 
Ice carried vertically by deep cumulus convection is strongly out of equilibrium with its environment, so that sublimation of convective ice produces a strong isotopic enhancement, exceeding the subsequent depletion when that added moisture is removed through re-deposition and precipitation of ice crystals.   
Processes with net zero effect on water vapor concentrations thus still leave an isotopic signature in residual vapor. Isotopic profiles are therefore uniquely suited for diagnosing the joint effects of sublimating convective ice followed by formation of in-situ cirrus.

Measured tropical profiles of HDO/H$_2$O generally show a ``turnaround'' in the TTL, from progressive isotopic depletion to enhancement  (Fig.\,\ref{fig:fig1}), that has been identified as a sign of convective ice sublimation \cite{Moyer-1996,Dessler-2003,Smith-2006,Sayres-2010,Randel-2012}. The robustness of this behavior is confirmed by the ACE solar-occultation instrument on the SCISAT-1 satellite, \cite{Bernath-2005,Nassar-2007} 
which provides a decade of isotopic profile retrievals with near-global coverage and effective vertical resolution of $\sim$2 km in the upper troposphere. (See SI S1).
In this work, we use relatively simple modeling to invert the ACE observations and obtain the first quantitative estimates of the sublimation rate of convective ice in the TTL.

\section*{Model}

Our bulk model represents the TTL with two equations, for the budgets of water and deuterated water, that  
account for the partially counteracting effects of deep convection.
Convective transport of \textit{ice} provides net moistening and isotopic enhancement, since isotopically heavy ice sublimates as overshoots mix with surrounding undersaturated air. 
Convective transport of \textit{vapor} provides net isotopic depletion, since convective vapor is lighter than its environment, and can moisten or dehydrate the environment depending on the temperature of the saturated detraining plume. If overshooting plumes rise adiabatically they are colder and drier than their surroundings, but thermal exchange and mixing can raise temperatures close to that of the environment \cite{Sherwood-2001}. We take as the basecase assumption that detraining plumes have reached environmental temperatures (and so are neutrally buoyant), but consider in sensitivity analysis cases as cold as adiabatic ascent (Fig.\,S1).

\begin{figure*}[ht!!!]
\centering
\includegraphics[width=0.9\textwidth]{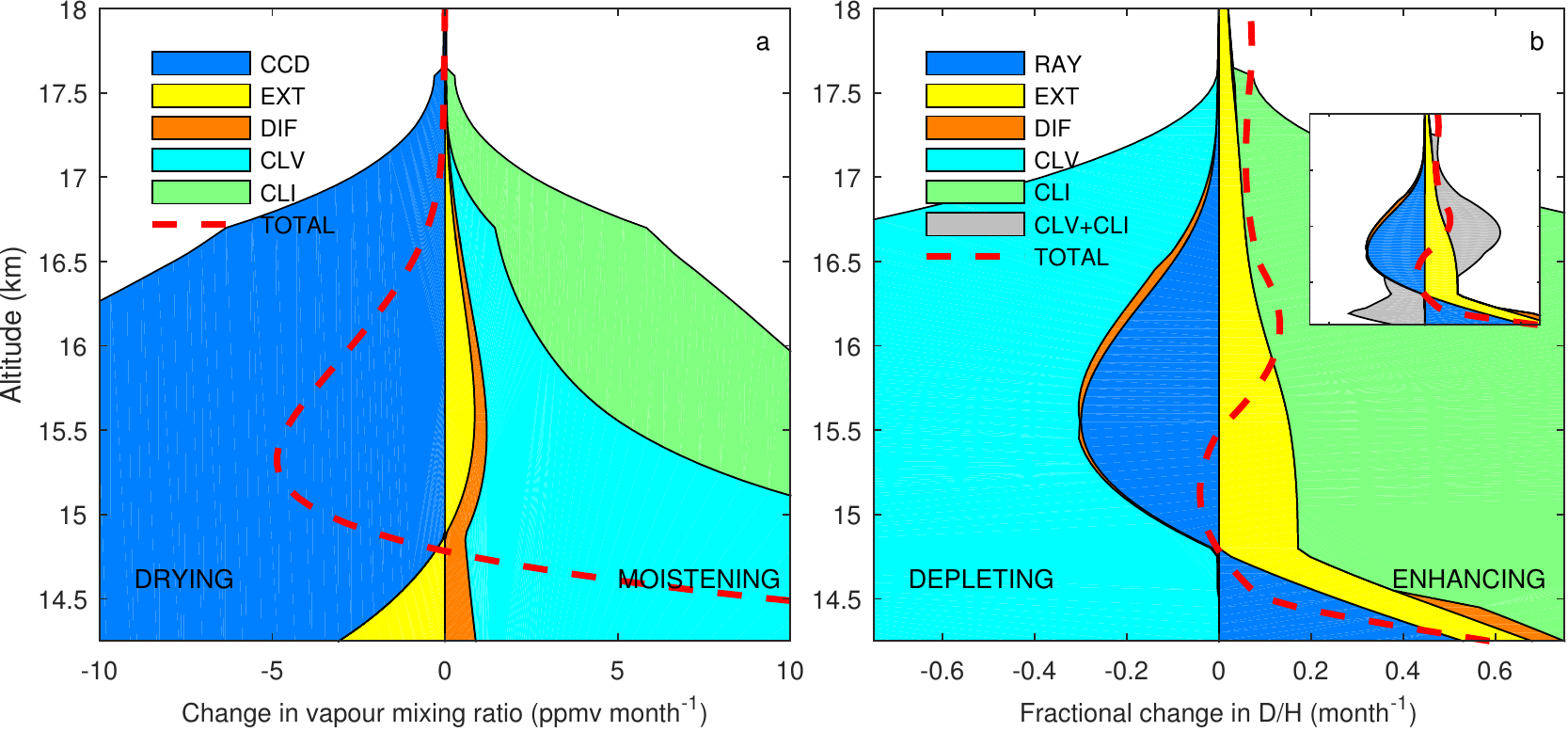} 
\caption{
Tendencies of TTL water vapor (a) and water isotope (b) budgets (i.e.\ moistening/drying rates and isotopic enhancement/depletion rates) from the model described in text, using basecase assumptions. Each budget shows five terms reflecting different processes. In both budgets, convective sources dominate. \textbf{a)} In the water budget (Eq.\ \ref{eq:rv}), CCD is cirrus condensation that removes vapor; water sources EXT and DIF are extratropical mixing and vertical diffusive transport, and CLV and CLI are vapor and sublimating ice from convective detrainment. CLV moistens in the basecase since air is assumed saturated at environmental temperature. Dashed line is the total net tendency on water, i.e.\ the change produced from the mean large-scale velocity alone. (Positive values occur where air is subsiding.)
\textbf{b)} In the isotope budget (Eq.\ \ref{eq:lnRv}), RAY is the depletion that would result simply from large-scale uplift; each other term captures the \textit{net} isotopic effect of water addition by a source followed by removal via cirrus formation.
Total net tendency on isotopic ratio (dashed line) is positive in the TTL above 15.5 km
because isotopic ratios are enhanced as air ascends, and positive below the TTL because isotopic ratios decrease with altitude but air is subsiding. 
Inset panel repeats budget with CLI and CLV summed to show net convective isotopic effects, which become positive in the TTL: sublimation of heavy convective ice outweighs the depletion from isotopically light convective vapor. \label{fig:fig2}}
\end{figure*}

We model the water budget in the TTL with a traditional ``leaky pipe'' model \cite{Mote-1998,Neu-1999} with added convection and cirrus formation processes: 
\begin{equation}
\small
w\partial_z r_v  = D\left(r_{vc}-r_v\right) + e
-c + K_{ex}\left(r_{vex}-r_v\right)+
K_v \partial^2_{zz} r_v
\label{eq:rv}
\end{equation}
\normalsize
where $r_v$ is the water vapor mixing ratio. The left-hand side is the vertical transport of water vapor by the large-scale vertical velocity $w$. On the right-hand side, the five terms are, in order, (i) moistening by detraining convective vapor with mixing ratio $r_{vc}$, (ii) moistening by sublimating convective ice $e$, (iii) loss of water by in-situ cirrus formation $c$, (iv)  mixing with extra-tropical air with mixing ratio $r_{vex}$, and (v) vertical diffusion. 
Several of these terms are constrained by observations;
we use ERA-Interim reanalysis \cite{Dee-2011} for vertical profiles of $w$, $r_v$ and $r_{vex}$ (though with a correction factor on $w$, see SI S2.5). 
We take estimates of the convective detrainment profile $D$ from \cite{Folkins-2006} (expressed as a rate of dilution of the TTL) and of the extra-tropical mixing rate $K_{ex}$ and diffusion coefficient $K_v$ from \cite{Mote-1998}. 
We assume that convective vapor $r_{vc}$ follows saturation at plume temperature. Eq.\,(\ref{eq:rv}) then provides an estimate of the net moisture source $e-c$, but cannot disentangle sublimation $e$ and condensation $c$. 
(See SI S2.1 for model derivation and Table S2.6 for default values of all model parameters.)

We resolve the degeneracy between sublimation $e$ and condensation $c$ by modeling the mean tropical isotopic profile.
Considering the isotopic effects of the same physical processes as in Eq.\,(\ref{eq:rv}) yields an 
equilibrium equation for the isotopic profile (expressed as the log of the isotopic ratio, $\ln R_v$) as:
\begin{equation}
{w}\partial_z \ln R_v =  \mathrm{RAY} + \mathrm{CLV} + \mathrm{CLI} + \mathrm{DIF} + \mathrm{EXT}\;
\label{eq:lnRv}
\end{equation}
\vspace{-0.03in}
with
\vspace{-0.00in}
\begin{align*}
& \mathrm{RAY}  = {}  \left(\alpha_i - 1\right) \partial_t \ln r_v \\
& \mathrm{CLV}  = {} D\left(-r_{vc}/r_v\left(\alpha_i - R_{vc}/R_v\right) + \left(\alpha_i-1\right)\right) \\
& \mathrm{CLI}  = {} -e/r_v\left(\alpha_i - R_{ic}/R_v\right) \\
& \mathrm{DIF}  = {} K_v\left(1/R_v \partial^2_{zz}R_v + 2 \partial_z \ln r_v \partial_z \ln R_v - \left(\alpha_i-1\right)/r_v \partial^2_{zz} r_v\right) \\
& \mathrm{EXT}  = {} K_{ex}\left(-r_{vex}/r_v\left(\alpha_i - R_{vex}/R_v\right) + \left(\alpha_i-1\right)\right)
\end{align*}
\normalsize
(See SI S2.2 for full derivation.) Here CLV, CLI, DIF, and EXT are the \textit{net} isotopic effects produced by the addition of water and its subsequent removal via cirrus formation, from four sources: convective vapor (CLV), 
sublimating convective ice (CLI), 
extratropical moisture (EXT), 
and vertical diffusion (DIF).
The fifth term  (RAY) is the effect that would result from large-scale uplift in the absence of any other water sources. This ``Rayleigh distillation'' describes the isotopic 
depletion that would be associated with
ice deposition producing the observed falloff of TTL water vapor with altitude.

Parameters and variables specific to the isotope budget include the isotopic fractionation factor during ice deposition ($\alpha_i$, taken from literature), vapor isotopic profiles in the tropics and extratropics ($R_v$ and $R_{vex}$, measured by ACE) and the isotopic profiles of deep convective water vapor $R_{vc}$ and ice $R_{ic}$. (Here $R_{ic}$ refers only to that ice that will sublimate).
Both $R_{vc}$ and $R_{ic}$ are uncertain, as no comprehensive observations exist of in-cloud isotopic compositions. We estimate $R_{vc}$ by assuming that convective vapor follows 
an isotope-resolving model of adiabatic ascent \cite{Bolot-2013}, with a small positive offset due to contamination by ice as the plume warms. We take $R_{ic}$ as a constant, somewhat lighter than in the adiabatic model to reflect preferential sublimation of smaller and more recently-formed ice crystals. The basecase assumptions are set for consistency with observed TTL ice loads (discussed in more detail later), but we vary both parameters widely in the sensitivity analysis.

\section*{Results}

The observed water vapor and isotopic profiles allow solving the paired budget equations (\ref{eq:rv}) and (\ref{eq:lnRv}) for $e$ and $c$; the resulting solution allows us to evaluate the relative importance of different sources of water and deuterated water to the TTL.  
Results for basecase assumptions 
imply that convection dominates the budgets of both water and water isotopes (Fig.\,\ref{fig:fig2}).  

\begin{figure}[t!!!]
\centering
\includegraphics[width=0.45\textwidth]{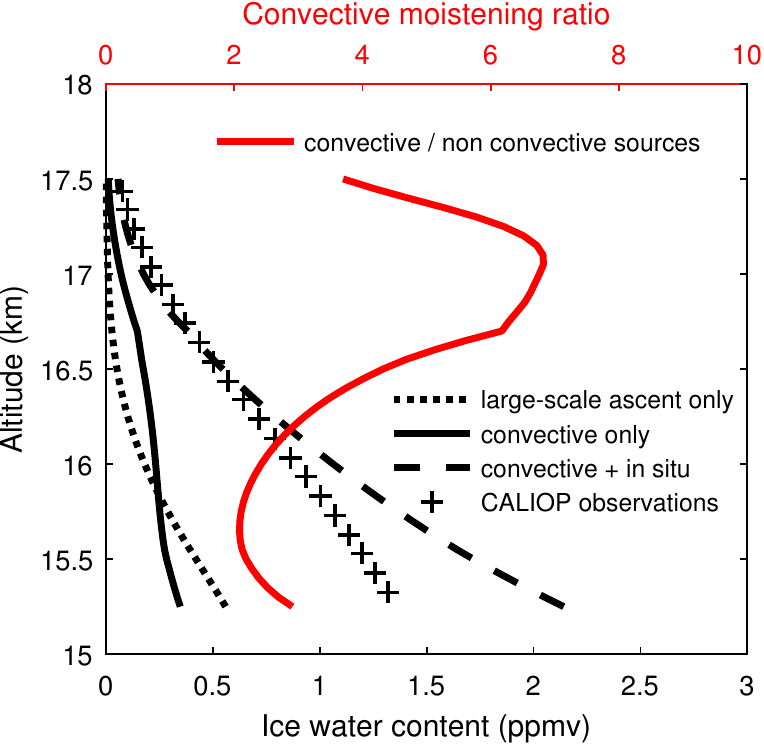}
\caption{Vertical profiles of ice water content suggested by moisture sources derived in this work (assuming sedimentation speed of 4 mm s$^{-1}$, sublimation timescale of 1 day for convective ice, and basecase parameter values). 
Solid line shows detrained convective ice alone; dashed line shows total cirrus, dominated by {\it in-situ} formation. 
Results are consistent with measured ice water content by CALIOP (crosses) and much higher than values expected from mean uplift alone (dotted line).  (Note that CALIOP may underestimate total ice loads in the lower TTL.)
Red line shows inferred ratio of convective sources to all non-convective sources (uplift, extratropical mixing and vertical diffusion).
\label{fig:fig3}}
\end{figure}

In the water budget (Fig.\,\ref{fig:fig2}a), the dominant TTL water sources are detraining convective vapor and sublimating convective ice.  
At $\sim$16.5 km, for example, convection provides 81\% of the total water source, with 66\% from convective ice and 15\% from convective vapor.
Deep convection therefore substantially amplifies the rate of TTL cirrus production over that expected from gradual uplift alone.
At $\sim$16.5 km, the rate of water removal by in-situ cirrus formation is nearly an order of magnitude larger than would occur with large-scale uplift alone. (Compare width of blue area in Fig.\,\ref{fig:fig2}a, $>$8 ppm/month, to dashed line, $\sim$0.9 ppm/month.)  Averaged over the whole TTL, convective ice and vapor together exceed all other sources combined by a factor of $\sim$2--6. (Red line in Fig.\,\ref{fig:fig3} shows ratio of convective to other sources.) 

In the isotope budget (Fig.\,\ref{fig:fig2}b), convective ice drives the TTL isotopic enhancement and produces the turnover in the isotopic profile. While in-mixed extratropical air does provide a source of heavier water, 
its isotopic effect is smaller than that of convection
and alone would be is too weak to counteract isotopic depletion during progressive drying of TTL air. (Compare yellow to blue areas in Fig.\,\ref{fig:fig2}b.)
Extratropical mixing without deep convection could produce the observed TTL isotopic profile only if rates were increased to unphysical values.  Mixing would have to dilute the TTL on timescales of $\sim$1 month, as compared to the $\sim$3 months estimated by \cite{Mote-1998}, our conservative basecase assumption, and over 12 months by others \cite{Volk-1996,Ploeger-2012}. (See SI 3.1.) The primary source of isotopic enhancement in the TTL appears to be lofted convective ice.

Note that the onset of isotopic enhancement at TTL altitudes 
occurs where detrainment rates are falling strongly (Fig.\,S2).
Enhancement results not from an increase in convective influence but from a change in the isotopic signature of convection, reflecting a transition in the balance of convective moisture sources.  
Below the TTL,  convective moisture deposited during detrainment is predominantly isotopically light convective vapor, and the net effect is isotopic depletion. In the TTL, convective moisture transport becomes dominated by isotopically heavy ice, and the net effect is enhancement. 
The transition occurs at $\sim$15.1 km in our basecase model, just above the level of zero net radiative heating at 14.7 km (Fig.\,\ref{fig:fig2}b inset).

Because all added water must be removed, 
moistening rates are directly related to {\it in-situ} cirrus production.
We therefore translate our derived moisture sources into estimated TTL ice water content, separately estimating ice loads for {\it in-situ} cirrus and the sublimating component of convective outflows. (We assume a sublimation timescale of 1 day and sedimentation velocity of 4 mm/s; see SI S2.3 for details.)
This calculation informs the basecase assumptions for the uncertain isotopic compositions of convective ice and vapor, which are chosen to produce ice loads similar to 
cirrus retrievals from the CALIOP (Cloud-Aerosol Lidar with Orthogonal Polarization) instrument, the primary instrument on board the CALIPSO satellite. CALIOP is sensitive to the small particles characteristic of {\it in-situ} cirrus and provides the only tropics-wide, vertically resolved measurements of TTL cirrus, though its measurements remain uncertain by a factor of $\sim$2 \cite{Avery-2012}.  
CALIOP observations (Fig.\,\ref{fig:fig3}, black crosses) imply a mean
ice load profile of $\sim$1.5--0.1 ppmv from base to top of the TTL; 
our model (Fig.\,\ref{fig:fig3}, dashed line) matches this falloff well in the upper TTL. 
If convective sources were disregarded in our model, inferred TTL ice water content would be lower by about a factor of five (Fig.\,\ref{fig:fig3}, dotted line). Those unrealistically low values are outside any plausible error for CALIOP and are similar to estimates of ice water content from TTL modeling studies that do not include convection \cite{Holton-2001,Dinh-2010}.

Given the uncertainty in many model parameters, as well as in observed TTL ice loads, we evaluate the robustness of these results
by conducting a sensitivity analysis over the most uncertain factors. 
We repeat the model solution described above 10,000 times, in each case sampling nine physical parameters  --
vertical velocity, detrainment rate, extratropical mixing rate, vertical diffusivity, ice crystal fall speed, in-cloud water vapor, and the isotopic compositions of convective vapor, convective ice, and extratropical vapor -- from distributions over their plausible ranges. (See SI S3 
for details.)  To understand the implication of uncertainties in ACE measurements or in isotope physics, we also repeat the basecase analysis using two different isotopic profiles (ACE version 3.5 and 2.2 retrievals), and three different estimates of the HDO/H$_2$O vapor-ice isotopic fractionation factor \cite{Merlivat-1967,Ellehoj-2013,Lamb-2016}.

\begin{figure*}[ht!]
\centering
\includegraphics[width=0.9\textwidth]{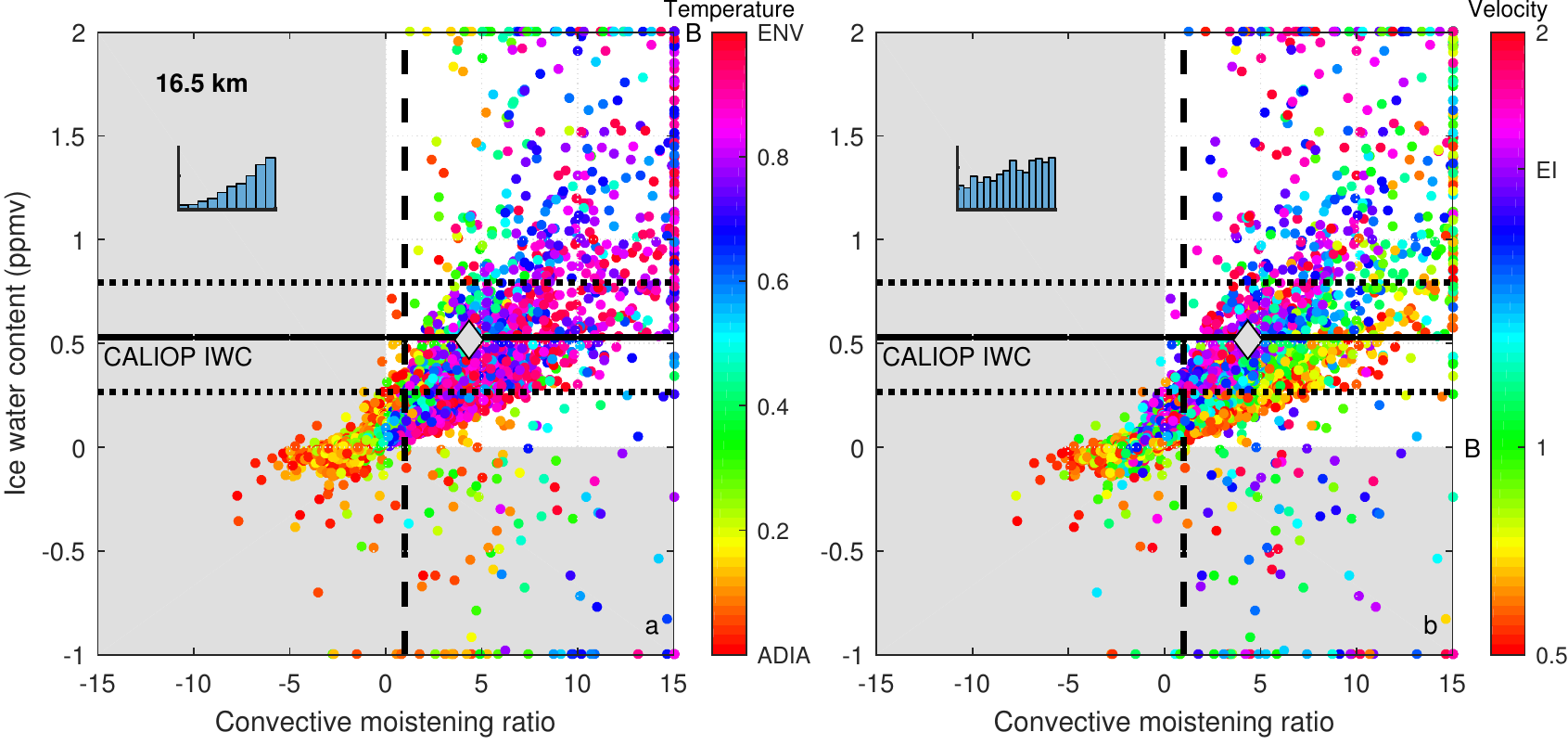} 
\caption{
All 10,000 solutions from the sensitivity analysis at 16.5 km, color coded by \textbf{a)} temperature in detraining plumes and \textbf{b)} vertical velocity, both as linear scaling indices. (SI Figs.\,S6-7 repeat for other variables.)
Outliers are shown on axes. 
`B's on each color scale mark basecase assumption; diamond marks basecase solution. On velocity scale, `EI' shows uncorrected velocity derived from ERA-Interim. 
Black lines show mean tropical ice load from CALIOP (solid) and $\pm$50\% of that value (dotted). Dashed line shows convective moistening ratio of 1, i.e.\ convective sources equal all other water sources. 98\% of solutions within dotted lines have convective moistening ratio $\ge 1$.
Grey regions mark solutions that are unphysical (negative ice water content) or that produce convective dehydration  (negative moistening ratio). Inset panels show histograms of the corresponding indices (temperature and velocity) for solutions producing ice loads $\pm$50\% from CALIOP.
\label{fig:fig4}}
\end{figure*}

The sensitivity analysis produces a family of solutions with the expected compact relationship between convective influence and TTL ice load. 
(See Fig.\,\ref{fig:fig4} and SI Figs.\,S6--10; each panel shows all 10,000 cases for 16.5 km altitude.)  
Solutions overwhelmingly imply that convection is a positive source of water to the TTL. Since TTL isotopic enhancement requires a substantial contribution from isotopically heavy convective ice, cases with net convective dehydration are almost always unphysical. 
Solutions also generally imply that convection is the dominant TTL water source. This conclusion is strengthened if ice loads are constrained to be consistent with CALIOP ($\pm$50\%): at 16.5 km,
98\% of these solutions show convective moistening ratios greater than 1. 
These results are robust to uncertainties in isotopic measurements and physics: substituting either the earlier ACE v2.2 retrieval or the substantially higher fractionation factor proposed by \cite{Ellehoj-2013} would increase the inferred convective contribution still further (Fig.\,S5). (Note that while using the value of \cite{Ellehoj-2013} would nearly double the implied convection contribution, recent results of \cite{Lamb-2016} are more consistent with basecase assumptions.)  

Finally, results consistently imply that convective moisture addition to the TTL is dominated by lofted ice. Even in the warmest detraining plumes, convective vapor never provides more than half as much moistening as does ice
 (Fig.\,S9j'). The water sources carried by deep convection have not been well known, and 
he proportion of lofted ice (that will sublimate) to water vapor inside convective towers 
has been an outstanding question in climate modeling. Our results suggest that this ratio is most likely in the range 1--3 (Fig.\,S9k).

Solutions show the greatest sensitivity to three parameters that are arguably the most uncertain of those tested: the temperature (and vapor content $r_{vc}$) of detraining plumes, the large-scale vertical velocity $w$ in the TTL, and the isotopic composition of sublimating convective ice $R_{ic}$.
We therefore discuss the implications of these analyses in more detail below. Fig.\,\ref{fig:fig4} left and right panels are color-coded by the first two of these, and SI Figs.\,S6--10 show all parameters.

For plume temperature, varying assumed conditions from purely adiabatic to purely environmental produces a spread in water vapor $r_{vc}$ of an order of magnitude in the mid-TTL (Fig.\,S1). 
In general, the warmer and therefore wetter the detraining plumes, the greater the inferred convective influence, not only because plumes carry more water vapor but because more convective ice is then required to counteract the isotopically depleted vapor and close the isotope budget. 
Cases with the coldest and driest detraining plumes generally cannot satisfy both water and isotope budgets and produce unphysical solutions 
with negative ice loads (Fig.\,\ref{fig:fig4}a). (This is especially true when mass fluxes from detrainment are large and those from large-scale ascent are small.) Purely adiabatic cases appear nearly impossible, and the distribution of solutions is skewed toward environmental temperatures. (Fig.\,\ref{fig:fig4}a inset shows a histogram of temperature indices for cases with ice within $\pm$50\% of CALIOP.)
Isotopic analysis can thus provide insight into convective dynamics in the TTL: it suggests that detraining convective plumes experience at least a moderate degree of mixing and warming. 

Vertical velocity in the TTL is not directly measurable and so is commonly estimated from the energy or momentum budget \cite{Abalos-2015}. Resulting estimates differ by factors of several depending on assumed trace gas concentrations and cloud radiative effects (Fig.\,S3 and Table S3.3; our velocity index range of 0.5--2 corresponds to ascent rates of 0.2--0.9 mm/s at 100 mb). The lower the assumed vertical velocity, the smaller the water source associated with large-scale uplift, and the greater the inferred convective influence relative to other sources, even if convective transport changes little. This factor produces most of the horizontal spread in the sensitivity analysis plots (Fig.\,\ref{fig:fig4}b). Isotopic modeling therefore only weakly informs estimates of TTL vertical velocities (Fig.\,\ref{fig:fig4}b inset). However, the exact value of vertical velocity becomes important for understanding the TTL water budget only if convective transport is relatively small. If the CALIOP ice load measurements are approximately correct, convection appears sufficiently dominant that large-scale uplift plays a comparatively minor role in bringing water to the TTL. 

The isotopic composition of sublimating TTL ice has not been directly measured. A single measurement in convective residue in the overworld stratosphere implies relatively heavy ice lofted from much lower altitudes (isotopic composition of -210$\pm$60\permil, i.e.\ only $\sim$20\% depleted in HDO relative to mean sea water) \cite{Hanisco-2007}. 
However, sublimation will likely be less extensive in the TTL than in the severely undersaturated stratosphere, with sublimating ice crystals likely smaller, more recent, and isotopically lighter.
We take as our basecase assumption an ice composition of -260\permil, but vary this value in sensitivity analysis from -500 to 0\permil.
The heavier the assumed ice composition, the lower the inferred convective influence, since less ice sublimation is then needed to produce a given enhancement (Fig.\,S6a). In our model, the heaviest ice values produce unreasonably low or even negative TTL ice loads. 
The lightest ice values, on the other hand, cause the model to break down as it loses ability to discriminate between sources: the outliers of Fig.\,\ref{fig:fig4} are all cases with ice composition near -500\permil. 
Ice composition can therefore be strongly informed by isotopic modeling. 
Fig.\,S8a shows the histogram of solutions at 16.5 km consistent with CALIOP; 
the mode of the distribution is indeed relatively light, at -450\permil. Inferred ice composition is correlated with inferred plume temperature, but the joint solution space consistent with CALIOP suggests that ice sublimating in TTL is isotopically lighter than that observed in the stratosphere (Fig. S10).

\section*{Discussion}
Despite wide uncertainty in individual parameters, our simple model and observed isotopic profiles, in conjunction with observed TTL cirrus ice loads, robustly imply that sublimating convective ice is the largest source of moisture to the TTL. 
Observed ice loads require large water sources, and isotopic profiles can only be matched if those sources are dominated by convective ice.  
Note however that ice crystals carried directly by convection make up only a small part of TTL cirrus in our model.  Since all added moisture must be removed by condensation and precipitation, convective sources also drive formation of  \textit{in-situ} cirrus  that is indirectly convective in origin,  
composed primarily of water recycled from sublimated convective ice. These ``secondary'' cirrus dominate ice loads in our model since their assumed lifetime against fallout exceeds that of primary convective ice against sublimation.  

Sensitivity analysis yields additional implications and suggests measurement priorities. Implications include that detraining convective plumes are warmed by mixing with the environment and that sublimating convective ice appears biased toward later-formed, isotopically lighter crystals.
Uncertainties would however be significantly narrowed if these two key parameters were known. Direct observations of detraining plume temperature and convective ice composition would allow drawing conclusions from isotopic profiles alone without the need for independent information on TTL ice loads.
 
If tropical high-altitude cirrus are driven by convective sources, then cirrus radiative impacts are tightly related to deep convection penetrating the TTL. Deep convection is expected to alter in future climate states (for example, some studies predict higher cloud top altitudes \cite{Zelinka-2010}), affecting the amount of primary ice lofted to the TTL and the production of secondary \textit{in-situ} cirrus. TTL cirrus may thus provide an important climate feedback.

\section*{Acknowledgements}
{The authors thank Eric Jensen, Zhiming Kuang, Kara Lamb, and Bill Randel for valuable comments and suggestions. 
The Atmospheric Chemistry Experiment (ACE), also known as SCISAT, is a Canadian-led mission mainly supported by the Canadian Space Agency. We also acknowledge support from the EU 7$^{th}$ framework Program under grant \#603557 (StratoClim), the NSF International Collaboration in Chemistry program, grant \#CHEM1026830, and the France-Chicago Foundation.}






\section*{Supplementary material}

# Supplementary online material
## Isotopic profiles imply strong convective influence on water near the tropical tropopause

Maximilien Bolot[1], Bernard Legras[2], Kaley A. Walker[3], Christopher D. Boone[4], Peter Bernath[5], William G. Read[6], and Elisabeth J. Moyer[1*]

[1]*Dept. of the Geophysical Sciences, University of Chicago, Chicago, IL, USA*
[2]*Laboratoire de Météorologie Dynamique, CNRS and Ecole Normale Supérieure, Paris, France*
[3]*Department of Physics, University of Toronto, Toronto, ON, Canada*
[4]*Department of Chemistry, University of Waterloo, Waterloo, ON, Canada*
[5]*Department of Chemistry and Biochemistry, Old Dominion University, Norfolk, VA, USA*
[6]*Jet Propulsion Laboratory, Pasadena, CA, USA*

# 1 Data

## 1.1 Water isotopic profiles

This analysis uses measurements of HDO and $H_2O$ from the ACE solar occultation Fourier transform spectrometer on board the SCISAT-1 satellite, launched in 2003 [1, 2]. We consider primarily the latest released version v3.5 (from 2013, [3]) but also compare to the prior version 2.2 (from 2005, [4]). The basecase analysis uses all v3.5 profiles over 9 years (2004-2013) that meet defined quality criteria and span all or part of the 13–18 km altitude region, a total of 700–1300 profiles in the tropics (15°S–15°N) and 5500–5900 in the extratropics (45–60°N/S) depending on altitude. Released data are interpolated to a 1 km vertical grid, but individual profiles have effective vertical resolution of ∼2 km in the TTL [5]. We average all qualifying measurements to obtain mean tropical and extra-tropical profiles (disregarding seasonal differences).

Quality criteria include that data flag 0 is checked; that estimated error in the isotopic composition (D/H ratio) is less than 200‰; and that values of D/H are within ±2× standard mean ocean water ratio (i.e. -2000–1000‰, large negative values are unphysical but must be retained in the average to avoid statistical bias). By averaging over a large number of profiles the residual standard deviation of the mean profile becomes negligible. Systematic errors linked to the retrieval procedure are a larger concern. We use the comparison between v3.5 and v2.2 as a proxy for the effect of unknown systematics. Systematics may be related to the sets of microwindows used for inversion of the measurements, which differ between retrieval versions, or to spectroscopic constants, which may be updated.

## 1.2 TTL ice water content

Total TTL ice water content measurements are derived from the Cloud-Aerosol Lidar with Orthogonal Polarization (CALIOP) lidar on board the CALIPSO satellite (part of the A-TRAIN constellation). CALIOP retrieves ice water content from cloud particle extinction profiles from scattered light at 532 nm and obtains over 20,000 profiles daily in the tropics (15°S–15°N). The data is gridded into 140 altitude bins from ∼12.5–19.5 km [6]. We use released version 3.01 IWC and average all tropical profiles that meet quality criteria from 3 June 2006 to 1 January 2011 (1558 days) into a single tropical mean profile. Profiles are rejected if the extinction quality flag is greater than 1 (indicating a poor retrieval), if the IWC uncertainty is stated as 99.99 (indicating a failed retrieval), or if the atmospheric volume description "MOD 8" is greater than 4 (which eliminates surface and opaque measurements). Note that CALIOP IWC is commonly assumed to be uncertain to a factor of 2. The v3.01 IWC profiles are similar to those derived from the Microwave Limb Sounder (MLS) instrument (±50%), but this agreement is coincidental: CALIOP is sensitive to small (several $\mu$m) ice particles of TTL in situ cirrus but insensitive to the large ice loads in the center of cumulus towers; MLS is sensitive only to large (>100 $\mu$m) convective ice particles and insensitive to TTL in situ cirrus [6].

## 1.3 Other variables

We derive profiles of temperature, radiative heating rates, large-scale vertical velocity and water vapor from the ERA-Interim reanalysis [7] (though note that we apply an adjustment factor to vertical velocities; see Section 2.5). We compute average tropical profiles using data from years 2004–2008 and latitudes 15°S–15°N, and a profile of extratropical water vapor using data from the same years and latitudes 45–60°N/S.

## 2 Models

### 2.1 Water vapor

The transport of non-condensing tracers in the TTL is usually represented according to the leaky tropical pipe model [8, 9]. This model applied to a tracer $\chi$ consists of a one dimensional advection-dilution-diffusion equation:

$$\underbrace{w\partial_z \chi}_{\text{advection}} = \underbrace{K_{ex}\left(\chi_{ex} - \chi\right)}_{\text{dilution}} + \underbrace{\frac{1}{\rho}\partial_z \left(\rho K_v \partial_z \chi\right)}_{\text{diffusion}}. \tag{S1}$$

where $\chi$ and $\chi_{ex}$ are the mixing ratios of the tracer in the tropics and midlatitudes; $w$ is the vertical velocity advecting tracers; $K_{ex}$ is the rate of dilution of the tropics by in-mixing of air from the midlatitudes; $K_v$ is the vertical diffusion coefficient; and $\rho$ is the density of air.

Our model for water vapor adds representation of a source of moisture from convection and a sink due to cirrus cloud formation:

$$w\partial_z r_v = \underbrace{D\left(r_{vc} - r_v\right) + e}_{\text{convective moistening}} \underbrace{-c}_{\text{cirrus formation}} + K_{ex}\left(r_{vex} - r_v\right) + \frac{1}{\rho}\partial_z \left(\rho K_v \partial_z r_v\right). \tag{S2}$$

Here $D$ is the rate of convective detrainment; $r_v$, $r_{vc}$ and $r_{vex}$ are the mixing ratios of vapor in the tropics, inside convective clouds and in the midlatitudes; $e$ is the rate of convective ice sublimation; and $c$ is the rate of condensation that forms in-situ cirrus. The detrainment term includes two sources of water: the direct detrainment of convective vapor and the sublimation of convective ice. Since overshooting cloud parcels are thought to come to rest and leave the cloud at neutral buoyancy, once their temperature is the same as the temperature in the environment, $r_{vc}$ is assumed at saturation with respect to environmental temperature. In these conditions, both vapor detrainment and ice sublimation constitute a source of vapor to the TTL. In the sensitivity analysis, however, we relax this assumption and consider the possibility that convection detrains after zero or partial buoyancy equilibration, in which case the detraining air is colder than the environment and $r_{vc} < r_v$ is possible, ie. direct vapor detrainment dehydrates the environment.

We obtain a more suitable form of the water vapor model by expanding the diffusion term on the R.H.S.:

$$\boxed{\tilde{w}\partial_z r_v = D\left(r_{vc} - r_v\right) + e - c + K_{ex}\left(r_{vex} - r_v\right) + K_v \partial_{zz}^2 r_v} \tag{S3}$$

Here, as in [9], we define a modified vertical velocity $\tilde{w}$ as:

$$\tilde{w} = w + \frac{K_v}{H}, \tag{S4}$$

where $H$, the scale height of the atmosphere (i.e. $\rho = \rho_0 \exp(-z/H)$), and $K_v$, the diffusion coefficient, have been assumed constant with altitude. The additional term $K_v/H$ is tiny relative to the mean advective vertical velocity $w$: maximal values of $w$ in the TTL are $8 \cdot 10^{-4}$ m s$^{-1}$, while the default value of $K_v/H$ is $3.5 \cdot 10^{-9}$ m s$^{-1}$. We therefore consider $w \approx \tilde{w}$ and ignore the distinction henceforth, and in the manuscript.

## 2.2 Isotopic ratio

We construct a model for deuterated water vapor analogous to that for non-deuterated water vapor:

$$w\partial_z r'_v = D\left(r'_{vc} - r'_v\right) + R_{ic}e - \alpha_i \frac{r'_v}{r_v} c + K_{ex}\left(r'_{vex} - r'_v\right) + K_v \partial^2_{zz} r'_v. \tag{S5}$$

Here $r'_v$, $r'_{vc}$ and $r'_{vex}$ are the deuterated water mixing ratios corresponding to $r_v$, $r_{vc}$ and $r_{vex}$; $R_{ic}$ is the isotopic ratio of sublimating ice; and $\alpha_i$ is the equilibrium isotopic fractionation factor for the vapor-ice transition.

This equation involves several isotopic assumptions. First, we assume that ice deposition occurs at thermodynamic equilibrium, and ignore any kinetic modifications to fractionation. (We do however later test for sensitivity of the results to the assumed values of $\alpha_i$.) Second, we assume that the ice formed by vapor deposition as in situ cirrus does not interact again with TTL vapor. That is, we ignore any isotopic effects that might result from the sublimation and reformation of precipitating ice crystals during their descent through the TTL. This assumption allows us to treat the condensation term in Eq. (S5) as a Rayleigh process.

Combining Eqs. (S3) and (S5) yields a model for the isotopic ratio of tropical vapor:

$$K_v \frac{\partial^2 R_v}{\partial z^2} + \left(2K_v \frac{\partial \ln r_v}{\partial z} - w\right) \frac{\partial R_v}{\partial z} + \left(-(\alpha_i - 1)\frac{c}{r_v} - D\frac{r_{vc}}{r_v} - \frac{e}{r_v} - K_{ex}\frac{r_{vex}}{r_v}\right) R_v$$
$$+ D\frac{r_{vc}}{r_v} R_{vc} + \frac{e}{r_v} R_{ic} + K_{ex}\frac{r_{vex}}{r_v} R_{vex} = 0 \tag{S6}$$

where $R_v$, $R_{vc}$ and $R_{vex}$ are the ratios of deuterated to non-deuterated vapor in the tropics, inside convective clouds and in the midlatitudes. We then replace $c$ by its definition from Eq. (S3) and rearrange to yield the version of the model that we use in this study, with the tendency on the log of the isotopic composition $\ln R_v$ expressed as the sum of five terms:

$$w\partial_z \ln R_v = \text{RAY} + \text{CLV} + \text{CLI} + \text{DIF} + \text{EXT} \tag{S7}$$
$$\text{RAY} = (\alpha_i - 1) w\partial_z \ln r_v$$
$$\text{CLV} = D\left(-r_{vc}/r_v \left(\alpha_i - R_{vc}/R_v\right) + (\alpha_i - 1)\right)$$
$$\text{CLI} = -e/r_v \left(\alpha_i - R_{ic}/R_v\right)$$
$$\text{DIF} = K_v \left(1/R_v \partial^2_{zz} R_v + 2\partial_z \ln r_v \partial_z \ln R_v - (\alpha_i - 1)/r_v \partial^2_{zz} r_v\right)$$
$$\text{EXT} = K_{ex} \left(-r_{vex}/r_v \left(\alpha_i - R_{vex}/R_v\right) + (\alpha_i - 1)\right)$$

RAY is the tendency corresponding to *net* loss of water vapor during ascent in the TTL, i.e. to the falloff with altitude in the water vapor profile. The other four terms all represent the isotopic effects of moisture addition followed by complete removal by in situ cirrus formation, so that there is an isotopic effect but no net effect on water vapor. CLV, CLI, DIF and EXT represent, respectively, convective vapor detrainment, convective ice sublimation, vertical diffusion, and mixing with the extra-tropics.

The rate of convective ice sublimation can thus be extracted directly from the observed tropical isotopic profile:

$$e = \frac{w\partial_z \ln R_v - \text{RAY} - \text{CLV} - \text{DIF} - \text{EXT}}{(-\alpha_i + R_{ic}/R_v)/r_v} \tag{S8}$$

## 2.3 TTL cirrus ice content

Using simple assumptions, we can convert the inferred rate of in situ cirrus formation to an estimate of ice water content in the TTL at any given time. Ice water in the TTL (mixing ratio $r_i$) is made of convective anvil cirrus (mixing ratio $r_{i1}$, "primarily generated" ice) and in situ formed cirrus (mixing ratio $r_{i2}$, "secondarily generated" ice):

$$r_i = r_{i1} + r_{i2}$$

Both quantities are expressed as mixing ratios with regard to the total TTL volume. Note that since our analysis provides information only on convective ice that contributes to moistening the TTL, we are disregarding here that part of convective ice that falls out without sublimating. This is equivalent to assuming that non-sublimating ice particles fall out of the TTL rapidly enough that they contribute negligibly to observed ice loads. The resulting quantity is an appropriate analogue to CALIOP observations, since CALIOP is relatively insensitive to ice in the dense centers of convective anvils.

Since $r_{i1}$ refers only to that part of convective ice that sublimates in the TTL, it can be estimated simply, because we have already solved for the rate of convective anvil cirrus sublimation $e$ via Eq. (S8), If we assume a single sublimation timescale $\tau$, then the profile of convective anvil ice (that will sublimate) is given by:

$$r_{i1} = e \cdot \tau \tag{S9}$$

In this work we use a default assumption of $\tau = 1$ day, as in Holton and Gettelman 2001 [10].

The sublimation of convective anvils provides moisture to the TTL that eventually re-condenses as in situ formed cirrus; we have previously estimated the condensation rate $c$ at each altitude. We assume that in situ cirrus at each altitude is maintained dynamically by a balance between this cirrus production and losses by vertical advection, sedimentation, and mixing out to the extratropics, with an additional term describing dilution via detrainment of convective anvil plumes (which contain only primary, not secondary cirrus). The budget of in situ cirrus can then be written as:

$$\underbrace{w\frac{\partial r_{i2}}{\partial z}}_{\text{advection}} = \underbrace{-Dr_{i2}}_{\substack{\text{dilution through}\\\text{detrainment}}} \underbrace{+c}_{\substack{\text{moisture}\\\text{recondensation}}} + \underbrace{\frac{1}{\rho}\frac{\partial}{\partial z}(\rho v_t r_{i2})}_{\text{sedimentation}} - \underbrace{K_{ex} r_{i2}}_{\substack{\text{mixing with}\\\text{extratropics}}} \tag{S10}$$

where $v_t$ is the (mass-weighted) terminal fall speed of ice crystals. We set $v_t$ to a conservative 4 mm s$^{-1}$ as in Holton and Gettelman 2001 [10], on the low end of likely values but still higher than the upward vertical advective velocity of TTL air. The relatively large fall speed means that we can drop the diffusive term since it becomes negligible to the transport of ice, which is dominated by fallout. (This continues to hold in sensitivity analysis since we adjust $v_t$ only upwards.)

We estimate the profile of $r_{i2}$ by rewriting the in situ formed cirrus budget of Eq. (S10) as:

$$(w - v_t)\frac{\partial r_{i2}}{\partial z} = c - \left(D + K_{ex} + \frac{v_t}{H}\right) r_{i2} \tag{S11}$$

and integrating Eq. (S11) downward. We begin integration at the highest level of convective influence (17.5 km in our basecase model), with the boundary condition set to the ice water content measured by CALIOP at that level. (All inferred profiles will therefore match CALIOP at their highest altitude.)

## 2.4 Convective vapor and ice

Because the properties of penetrating convection in the TTL are relatively uncertain, we vary them widely in sensitivity analysis. Figure S1 shows profiles of assumed *(left)* mixing ratio of vapor in detraining plumes and *(right)* isotopic compositions of convective vapor and ice, for both the basecase and the range used in sensitivity analysis.

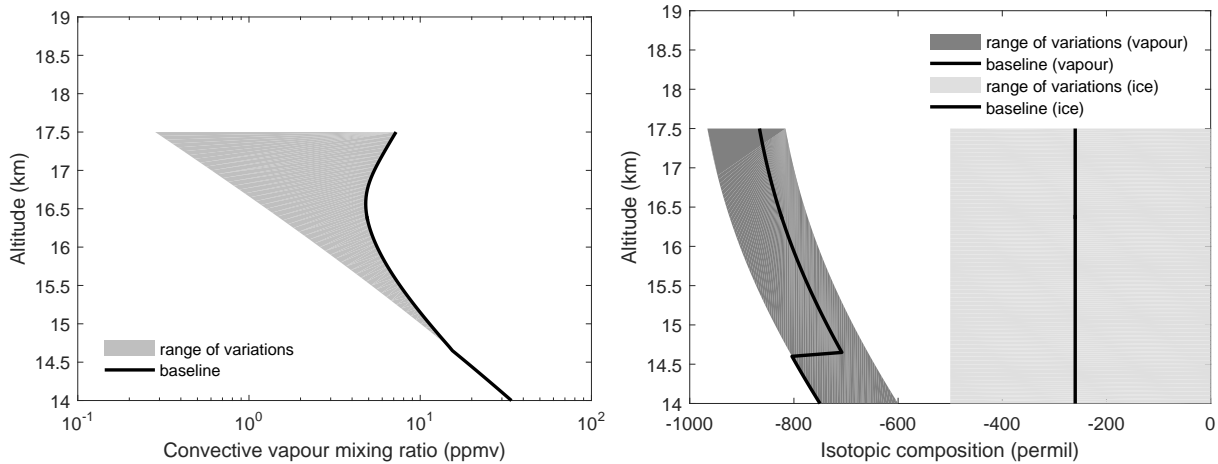

Figure S1: (Left): Profiles of the isotopic composition of convective vapor and sublimating ice in the base model (baseline) and their corresponding range of variation (vapor: dark grey; ice: light grey) used in the sensitivity analysis. (Right): Profile of convective vapor mixing ratio in the base model (saturation at environmental temperature) and its corresponding range of variation used in the sensitivity analysis (saturation at temperatures from adiabatic to environmental).

**Vapor:**

We assume that vapor in detraining plumes is saturated with respect to ice:

$$r_{vc} = r_{\text{sat}}^i [T].  \quad (S12)$$

This is equivalent to assuming that any supersaturation present in the updraft has been relaxed during the buoyancy adjustment that leads to detrainment. In the basecase model we assume that the temperature has warmed to environmental values, i.e. that $T$ is the ambient environmental temperature and that overshooting parcels detrain at neutral buoyancy. This "buoyancy equilibration" is supported by several studies (e.g. [11, 12]), but the processes involved in overshooting convection are still poorly known, so we consider a full range of lower temperatures in sensitivity analysis, down to that predicted by purely adiabatic ascent. In the most extreme scenario, parcels detraining at the top of the TTL can be fifteen degrees colder than the environment in which they detrain.

In the sensitivity analysis, we describe the convective vapor mixing ratio $r_{vc}$ as given by saturation at a modified temperature set by a control parameter $x$ that can vary from 0 (adiabatic) to 1 (environmental):

$$r_{vc} = r_{\text{sat}}^i [(1-x)T_c + xT] \quad \text{for} \quad z > \text{LNB}, \quad (S13)$$

where $T_c$ and $T$ are the adiabatic and environmental temperatures respectively. Setting $x$ to 0 reduces convective water vapor by over an order of magnitude over the basecase at the top of the TTL.

We assume that the isotopic composition of convective vapor falls off with altitude, but that it is somewhat enhanced over the values produced in adiabatic ascent. (Enhancement can occur because warming of overshooting plumes causes some sublimation of convective ice, or because glaciation is delayed, reducing the effective fractionation that an air parcel experiences.) We therefore represent convective vapor isotopic composition with the adiabatic updraft model of Bolot et al. 2013[1] [13] plus a constant offset. The basecase offset is set to 100‰; in sensitivity analysis we vary it from 0–150‰.

---

[1] The updraft model describes transformations of bulk vapor, liquid water and ice during adiabatic ascent, and is run with control parameters set as follows: saturation parameter = 0, glaciation parameter = 1.5, auto-conversion coefficient = 0.3, no Wegener-Bergeron-Findeisen effect, cloud base isotopic composition = -80 ‰. These settings mean that vapor pressure is saturated over ice in the glaciated part of the cloud, that glaciation roughly happens at -16°C and that liquid water is 50% depleted with respect to its adiabatic content at the onset of glaciation.

5

**Ice:**

We set the basecase isotopic composition of ice sublimating in the TTL to a constant value of -260‰ and vary it in sensitivity analysis from -500 to 0‰. The ice isotopic composition is highly uncertain, and this uncertainty propagates into the remainder of the analysis. The relatively light basecase assumption reflects the assumption that the crystals that sublimate in or around overshoots likely correspond to the smaller end of the size distribution [14] and so will be more recently formed and isotopically lighter than the bulk ice in the updraft. We assume that the larger particles that comprise most of the ice load sink back within convective clouds and are not detrained.

## 2.5 Vertical velocity, detrainment rate

Retrieving the ice sublimation rate from Eq. (S8) requires estimates of the large-scale vertical velocity $w$ and the convective detrainment rate $D$ in the TTL. Values used in the base model are shown in Figure S2 and the corresponding range of variations used in the sensitivity analysis in Figure S3.

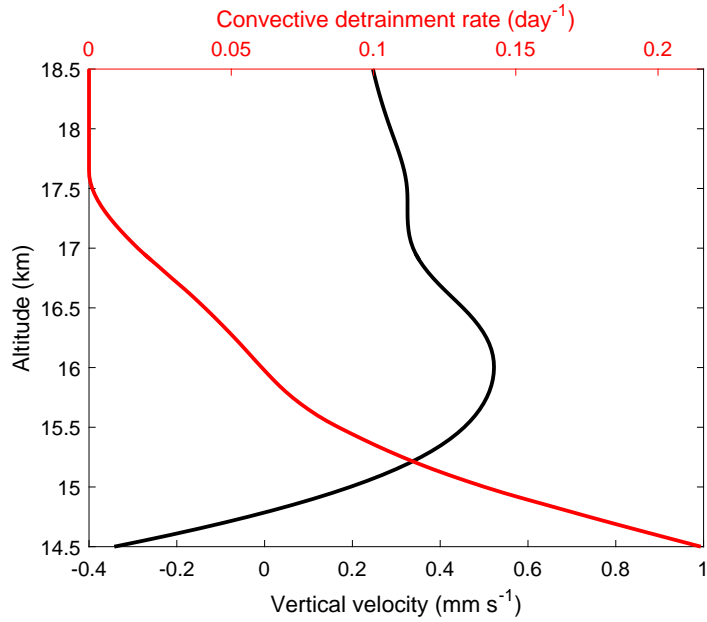

Figure S2: Vertical profiles of the vertical velocity $w$ and the convective detrainment rate $D$ used in the base model. The transition from subsidence to uplift occurs at 14.7 km.

**Vertical velocity:**

We assume, as in many previous studies (see e.g. [15]), that the large-scale vertical velocity $w$ is driven by the net radiative heating rate $Q_{\rm rad}$ (shortwave + longwave), and neglect other contributions to the TTL energy budget:

$$w\frac{\partial s}{\partial z} = Q_{\rm rad}, \tag{S14}$$

where $s$ is dry static energy ($s = c_p T + gz$) and $Q_{\rm rad}$ is the mean radiative heating rate in the TTL. (Other terms are as in Sect. 2.1.) Although we use ERA-Interim reanalysis [7] for other environmental profiles, several authors have found ERA-Interim to overestimate the diabatic velocities in clear sky conditions [16, 17]. We therefore derive the basecase velocity assumption by computing $Q_{\rm rad}$ from ERA-Interim (we take tropical averages over 15°S–15°N), but following Ploeger et al. 2012 [16], we apply a correction factor of 0.6.

For the sensitivity analysis, we consider a range of velocities (Fig. S3, left) from half to twice as large as the basecase profile (i.e. we apply a scaling factor ranging from 0.5 to 2). The resulting velocity range at the top of the TTL (100 hPa) is 0.2–0.9 mm s$^{-1}$, larger than the range estimated by Abalos et al. 2015 [18] (0.25–0.6 mm s$^{-1}$). Note that our range of variation also includes the uncorrected value of the diabatic vertical velocity estimated from ERA-Interim.

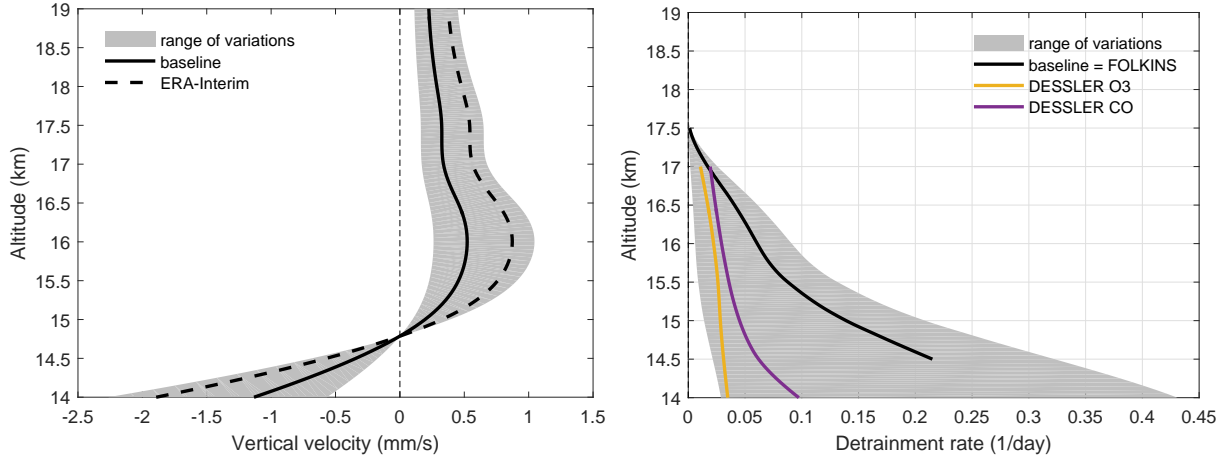

Figure S3: (Left): Profile of vertical velocity in the base model (baseline, derived from ERA-Interim radiative heating rate in the tropics, with a correction factor 0.6 [16]) and corresponding range of variations considered in the sensitivity analysis. (Right): Profile of detrainment rate in the basecase model and the range of variations considered in sensitivity analysis. For comparison we show estimates derived from considering measured tropical CO and O$_3$ profiles [19].

**Detrainment rate:**

We take the basecase detrainment rate $D$ from Folkins et al. 2006 [20], but also apply a correction discussed in their article that increases $D$ somewhat. (Compare Figure 1 of [20] with Figure S2.) The correction adjusts for the rate of mass divergence to the extratropics and is evaluated as $-1/\rho\, \partial\left(\rho\bar{w}\right)/\partial z$ with variables derived from ERA-Interim and taken as tropical averages from 15°S–15°N. $\bar{w}$ here is the reanalysis vertical velocity profile (which includes convective transport).

Because many other published estimates of detrainment rates are much lower than those of Folkins et al. 2006, we take an asymmetric range of variation for $D$ in our sensitivity analysis, multiplying the basecase value by a scaling factor ranging from 0.1 to 1.5 (Fig. S3, right). The resulting range encompasses the low detrainment values estimated by Dessler 2002 [19].

## 2.6 Data and assumptions used in this analysis

Table S2.6 below contains information on all variables directly or indirectly used in evaluating the rate of convective ice sublimation, Eq. (S8), and in estimating the resultant TTL cirrus ice load, Eqs. (S9) and (S11).

| # | Requires | Source | Details |
|---|---|---|---|
| $T$ | | ERA-Interim [7] | Temperature. ERA-Interim, averaged over 15°S-15°N from 2004–2008 |
| $T_c$ | | *updraft model* [13] | Temperature in convective clouds. Assumes adiabaticity |
| $s$ | $T$ | *derived* | Dry static energy |
| $Q_{\rm rad}$ | | ERA-Interim [7] | Radiative heating rate. ERA-Interim, averaged over 15°S-15°N from 2004–2008. Computed from clear sky longwave and shortwave tendency accumulation in the forecast model (ECMWF table 162, variables 102 and 103) |
| $\bar{w}$ | | ERA-Interim [7] | Vertical advective velocity inclusive of cloud transport. ERA-Interim vertical velocity, averaged over 15°S-15°N from 2004–2008 |
| $w$ | $s, Q_{\rm rad}$ | *derived* | Vertical advective velocity (Fig. S2) |
| $D$ | $\bar{w}$ | Folkins et al. 2006 [20] | Convective detrainment rate (Fig. S2). Includes correction for the rate of mass divergence to the extra-tropics |
| $K_{ex}$ | | Mote et al. 1998 [9] | Dilution rate from mixing with midlatitudes (Fig. S4) |
| $K_v$ | | Mote et al. 1998 [9] | Vertical diffusion coefficient $K_v = 0.03$ m$^2$s$^{-1}$ |
| $H$ | | *assumed* | Scale height $H = 8500$ km |
| $r_v$ | | ERA-Interim [7] | Vapor mixing ratio. Averaged over 15°S-15°N from 2004–2008 |
| $r_{vex}$ | | ERA-Interim [7] | Vapor mixing ratio in the extra-tropics. Averaged over 45°-60°N/S from 2004–2008, interpolated to tropical $\theta$ |
| $x$ | | *assumed* | Overshoot temperature index. Default $x = 1$, i.e. environmental temperature at detrainment |
| $r_{vc}$ | $T_c, T, x$ | *updraft model* [13] | Vapor mixing ratio in convective clouds. Computed as in Bolot et al. 2013 [13] up to neutral buoyancy level, then as $r_{\rm sat}^{\rm i}\left[(1-x)T_c + xT\right]$. (Fig. S1). |
| $R_v$ | | ACE-FTS [21] | Vapor isotopic ratio in tropics. From ACE-FTS v3.5 occultations (Feb 2004 to Jun 2013) meeting quality criteria defined in S1.1, averaged over 15°S-15°N, 700–1300 occultations |
| $R_{vex}$ | | ACE-FTS [21] | Vapor isotopic ratio in the extra-tropics. As above, but averaged over 45°N-60°N/S, 5500–5900 occultations |
| $\alpha_i$ | $T$ | Merlivat & Nief 1967 [22] | Ice-vapor fractionation coefficient. Equilibrium fractionation, no kinetic effects |
| $R_{vc}$ | | *updraft model* [13] + *assumed offset* | Vapor isotopic ratio in convective clouds. Computed from Bolot et al. 2013 [13] with control parameters as described in S2.4. + constant compositional offset (100‰ in default model) |
| $R_{ic}$ | | *assumed* | Isotopic ratio of convective sublimating ice. Assumed constant. Default -260‰ |
| $\tau$ | | Holton & Gettleman 2001 [10] | Timescale for sublimation of convective ice. Assumed constant. Default 1 day |
| $v_t$ | | Holton & Gettleman 2001 [10] | Sedimentation speed for in-situ ice. Assumed constant. Default 4 mm s$^{-1}$ |

Table S2.6: All variables directly or indirectly used in this analysis, with basecase values and sources.

# 3 Sensitivity analyses

## 3.1 Extratropical mixing rather than convection as source of TTL enhancement

To evaluate whether the isotopic 'turnover' in the TTL might be due *only* to mixing in of extra-tropical air with isotopically heavier water vapor, we evaluate the value of the dilution rate $K_{ex}$ that would allow closing the deuterium budget without convection, and consider whether it is realistic. Without convection, the isotopic budget (Eq. S7) simplifies to

$$w\partial_z \ln R_v = \text{RAY} + \text{DIF} + \text{EXT} \tag{S15}$$

We solve for $K_{ex}$ by expanding the term corresponding to mixing from the extra-tropics, yielding:

$$K_{ex} = \frac{w\partial_z \ln R_v - \text{RAY} - \text{DIF}}{(-r_{vex}/r_v\,(\alpha_i - R_{vex}/R_v) + (\alpha_i - 1))} \tag{S16}$$

We conservatively use for $r_{vex}$ and $R_{vex}$ the characteristics of air observed between 45°–60° latitude, to represent air with no influence of the dry subtropics. (Air at these latitudes is up to 40% wetter and 50-100‰ isotopically heavier than in the TTL). Results are displayed in Fig. S4 (as the timescale $1/K_{ex}$), along with the estimates of Mote et al. 1998 [9], the lower end of published estimates for the mixing timescale. If observed isotopic enrichment in the TTL were to be explained only by transport of isotopically heavier midlatitudes vapor, mixing timescales would have to as low as <1 month in the mid-TTL, consistently about half the Mote et al. values and far below other estimates: for example, Volk et al. 1996 [23] give an annual mean timescale of 13.5 months for $\theta$ in the TTL and lowermost stratosphere (from 370–420 K); Ploeger et al. 2012 [16] reevaluate that value to be 12.7 months. In the annual mean, [16] find that ∼10% of TTL air is in-mixed from the extratropics.

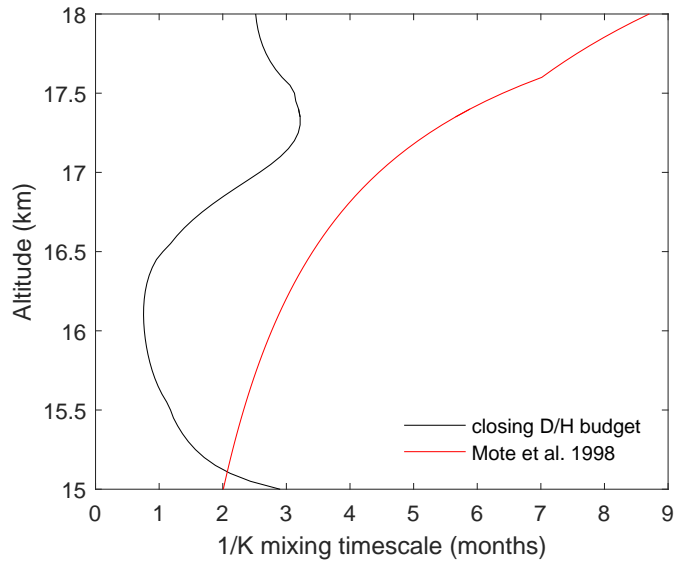

Figure S4: Vertical profile of the extratropical mixing dilution timescale ($1/K_{ex}$) required to close the deuterium budget, taking the characteristics of air mixing into the tropics as those of the 45–60° N/S latitude bands (solid black). The dilution timescale of Mote et al. 1998 [9] is plotted for comparison.

## 3.2 Robustness to assumed fractionation factor and ACE retrieval version

We separately test the sensitivity of the analysis to two sources of uncertainty, the value of the vapor-ice fractionation factor and the ACE measurements themselves. For the fractionation factor, the original measurement of Merlivat and Nief 1967 [22] used in this study is the most commonly assumed in the literature, but is poorly validated. In 2013 Ellehöj et al. suggested a substantial upward revision, with the preferential partitioning $\alpha_{eq}$-1 nearly 50% higher [24]; recent measurements by Lamb et al. [25] suggest instead a slight downward revision. The ACE measurements are sufficiently well-sampled for the precision error of the mean profile to be fairly small, but satellite retrievals are subject to unknown systematics. We assess the uncertainty due to systematic measurement error at least roughly by comparing results obtained with two versions of the ACE retrievals (v3.5 and v2.2), using the same occultations for both. Version 2.2 produces a slightly isotopically heavier profile with a lower turnaround point (see manuscript Fig. 1).

These revisions tend to reinforce the conclusion that convection dominates as source of water in the TTL. Figure S5 shows derived ice water content and convection moistening ratio given different values of the fractionation factor and different ACE retrieval versions. Using either the heavier v2.2 profile or the larger fractionation factor of Ellehöj et al. both increase the inferred convective moistening over the base case. (A larger fractionation factor produces isotopically lighter vapor, requiring more convective influence to produce a given enhancement.) Using the Lamb et al. fractionation factor does decrease inferred convective influence but only slightly, since the adjustment to Merlivat and Nief is small.

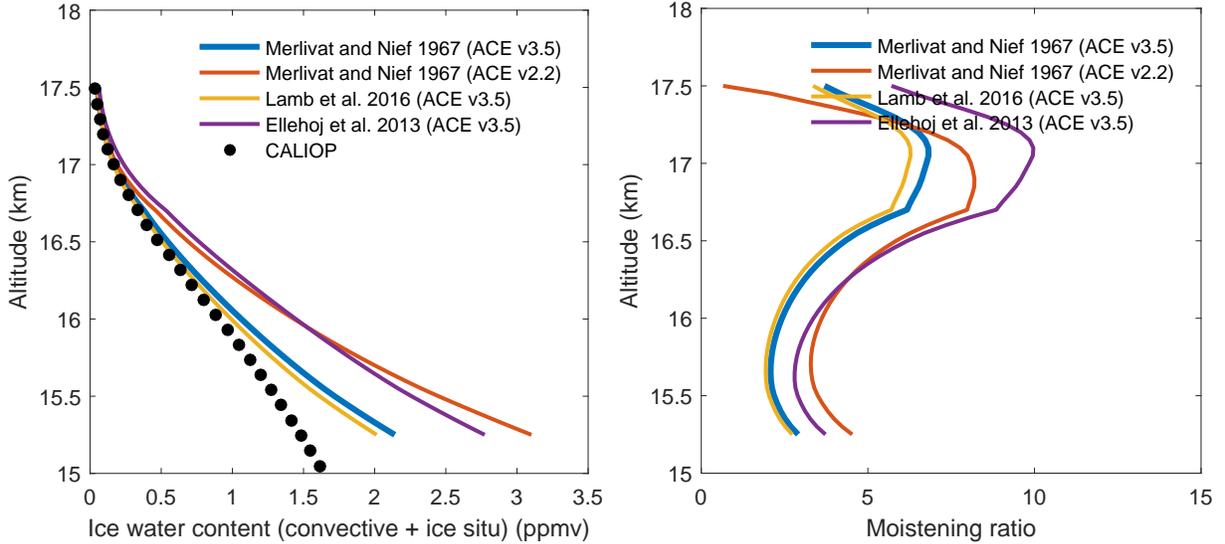

Figure S5: Profiles of derived TTL cirrus ice water content (left) and convective moistening ratio (right) under different ice-vapor fractionation factor values and ACE retrieval versions. Left panel shows measured TTL cirrus ice from CALIOP for comparison. Higher fractionation factors and heavier isotopic profiles (ACE v2.2) produce larger inferred convective influence.

10

## 3.3 Sensitivity to model parameter values

In the formal sensitivity analysis, we modify each of the nine parameters in Table S3.3 below. The analysis consists of 10,000 draws from assumed distributions of these variables. (We assume for each variable a uniform distribution between lower and upper bounds.) We assume that the variables are independently distributed and so perturb them individually; this assumption may not be fully true for e.g. water vapor content ($r_{vc}$) and isotopic composition ($R_{vc}$) in convective overshoots. The analysis does however cover what we consider the full range of plausible values and so covers all parameter space.

| # | Parameter adjustment (random variable underlined) | Lower | Upper | Default |
|---|---|---|---|---|
| $v_t$ | $\underline{v_t}$ | 3 mm s$^{-1}$ | 50 mm s$^{-1}$ | 4 mm s$^{-1}$ |
| | | \multicolumn{3}{l}{Yields: terminal velocities for ice effective diameters from $\sim$ 8 to 40 $\mu$m (range measured during ATTREX) [26, 27]} | | |
| $D$ | $D^0 \times \underline{f}$ | 0.1 | 1.5 | 1 |
| | | \multicolumn{3}{l}{Yields: detrainment timescales from 20 days to 11 months at 16.5 km} | | |
| $K_{ex}$ | $K_{ex}^0 \times \underline{f}$ | 0.5 | 1.5 | 1 |
| | | \multicolumn{3}{l}{Yields: mixing timescales from 2 to 7 months at 16.5 km} | | |
| $K_v$ | $K_v^0 \times \underline{f}$ | 0 | 2 | 1 |
| | | \multicolumn{3}{l}{Yields: from no diffusion to $K_v = .006$ m$^2$s$^{-1}$} | | |
| $r_{vc}$ | $r_{\text{sat}}^{\text{i}}\left[(1-\underline{x})T_c + \underline{x}T\right]$ | 0 | 1 | 1 |
| | | \multicolumn{3}{l}{Yields: from adiabatic to environmental} | | |
| $R_{vc}$ ($\delta_{vc}$ in $\delta$-notation[2]) | $\delta_{vc}^0 + \underline{\varepsilon}$ | 0‰ | 150‰ | 100‰ |
| | | \multicolumn{3}{l}{Yields: from -940 to -790‰ at 16.5 km (adds up to 150‰ to adiabatic)} | | |
| $R_{ic}$ ($\delta_{ic}$ in $\delta$-notation) | $\underline{\delta_{ic}}$ | -500‰ | 0‰ | -260‰ |
| $R_{vex}$ ($\delta_{ext}$ in $\delta$-notation) | $\delta_{ext}^0 + \underline{\varepsilon}$ | -50‰ | 50‰ | 0‰ |
| | | \multicolumn{3}{l}{Yields: from -640 to -540‰ at 16.5 km} | | |
| $w$ | $w^0 \times \underline{f}$ | 0.5 | 2 | 1 |
| | $w^0$: (ERA-Int.) $0.6 \times Q_{\text{rad}}/(\partial s/\partial z)$ | \multicolumn{3}{l}{Yields: vertical velocities from 0.2 to 0.9 mm s$^{-1}$ at 100 hPa} | | |

Table S3.3: Control parameters used in the sensitivity analysis and range of variation. Default values refer to the basecase.

We show results of the sensitivity analysis below (Figures S6–S9), with separate plots color-coded by each of the nine parameters and a tenth and eleventh parameter measuring the ratio of ice to vapor contributions in convective moistening ($e/(D(r_{vc} - r_v))$) and the ratio of ice to vapor abundance at the top of convective towers ($e/(Dr_{vc})$).

11

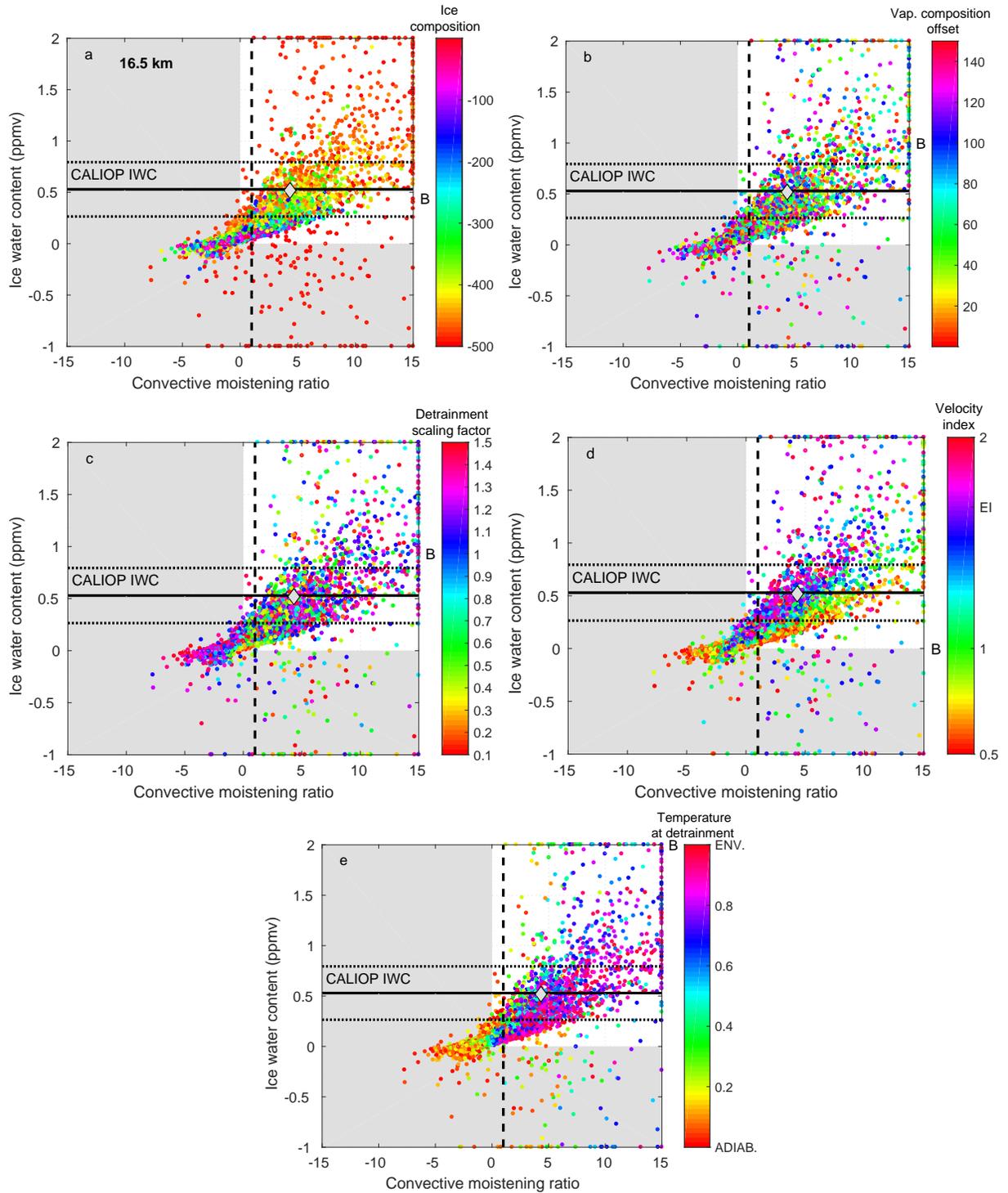

Figure S6: Inferred ice water content vs convective moistening ratio (convection/other sources) at 16.5 km for all 10,000 cases of the sensitivity analysis, color coded by (a): sublimating ice composition; (b): convective vapor isotopic composition offset; (c): scaling factor on the detrainment rate; (d): scaling factor on the vertical velocity; (e): index for the temperature at detrainment (linear from adiabatic to environmental). The symbol (B) on each color scale indicates the baseline assumption. (In (d), EI marks unadjusted ERA-Interim velocity.) Solid line shows CALIOP tropical IWC at 16.5 km, dashed lines show values within +/- 50%. Vertical dashed line marks convective moistening ratio 1. Areas in grey correspond to inferred negative ice water content that is physically impossible or negative moistening ratio (encompassing occurrences of convective dehydration). White diamond shows basecase solution. CONTINUES on Fig. S7 for color-coding by additional variables.

12

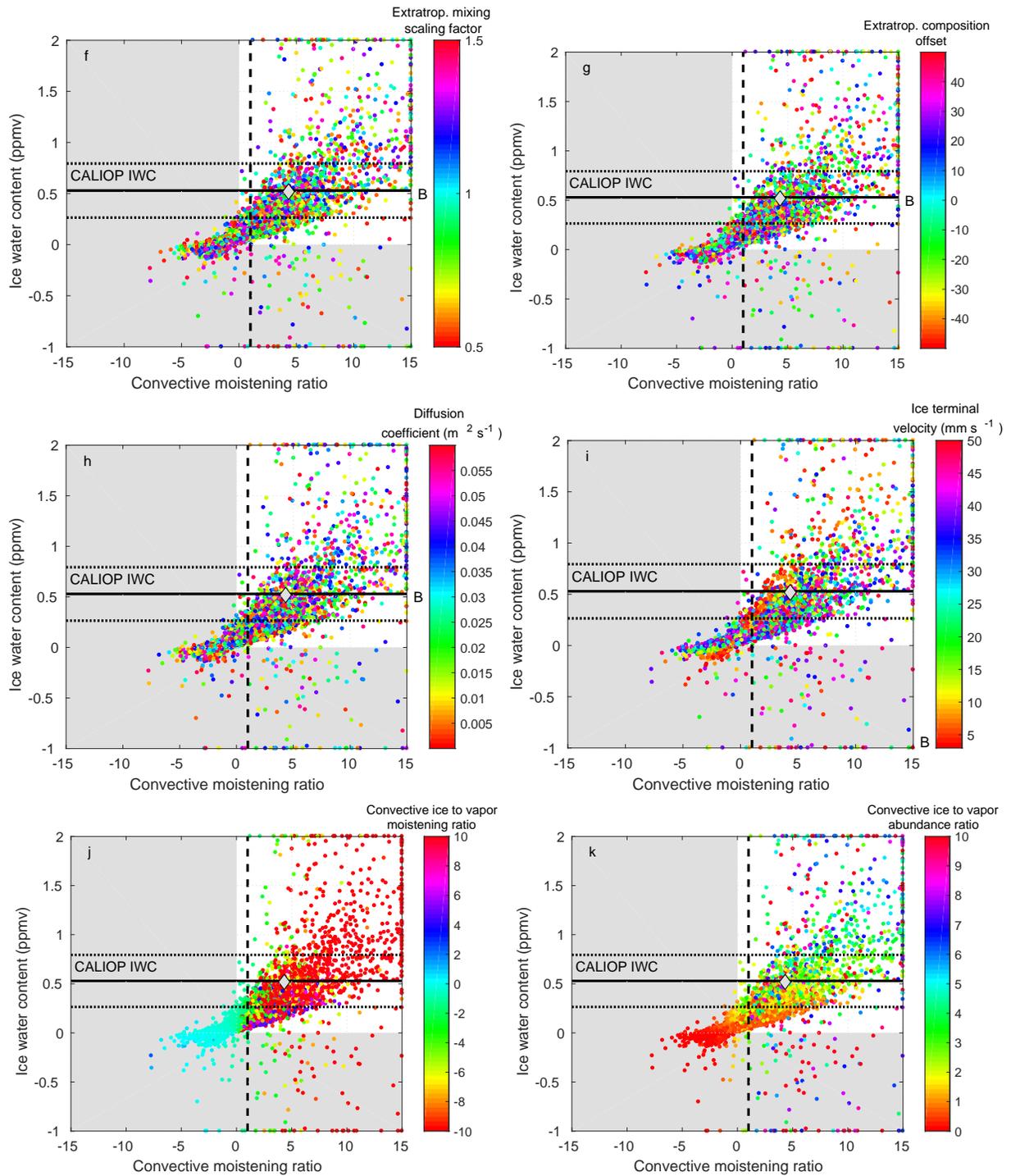

Figure S7: CONTINUED from Fig. S6, here color-coded by (f): scaling factor on extratropical mixing rate; (g): extratropical vapor isotopic composition offset; (h): value of the diffusion coefficient; (i): value of the ice terminal velocity; (j): ratio of ice to vapor contributions in convective moistening; (k): ratio of ice to vapor abundance at the top of convective towers.

13

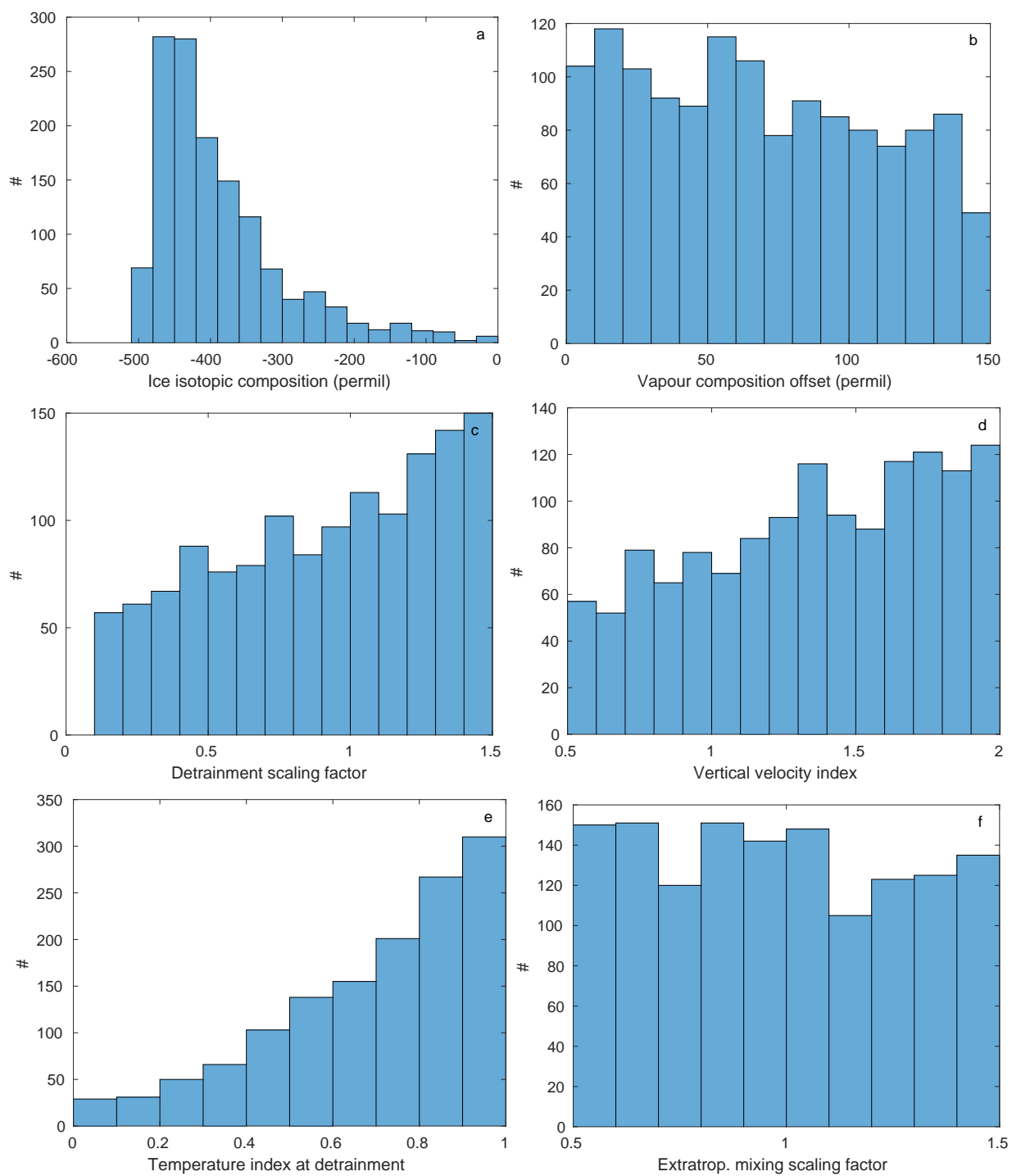

Figure S8: Histograms of model sensitivity analysis solutions at 16.5 km for (a): sublimating ice composition; (b): convective vapor isotopic composition offset; (c): scaling factor on the detrainment rate; (d): scaling factor on the vertical velocity; (e): linear index on the temperature scale at detrainment from adiabatic to environmental; (f): scaling factor on extratropical mixing rate, for those cases that yield ice within +/- 50% of CALIOP at this altitude. CONTINUES on S9 for additional variables.

14

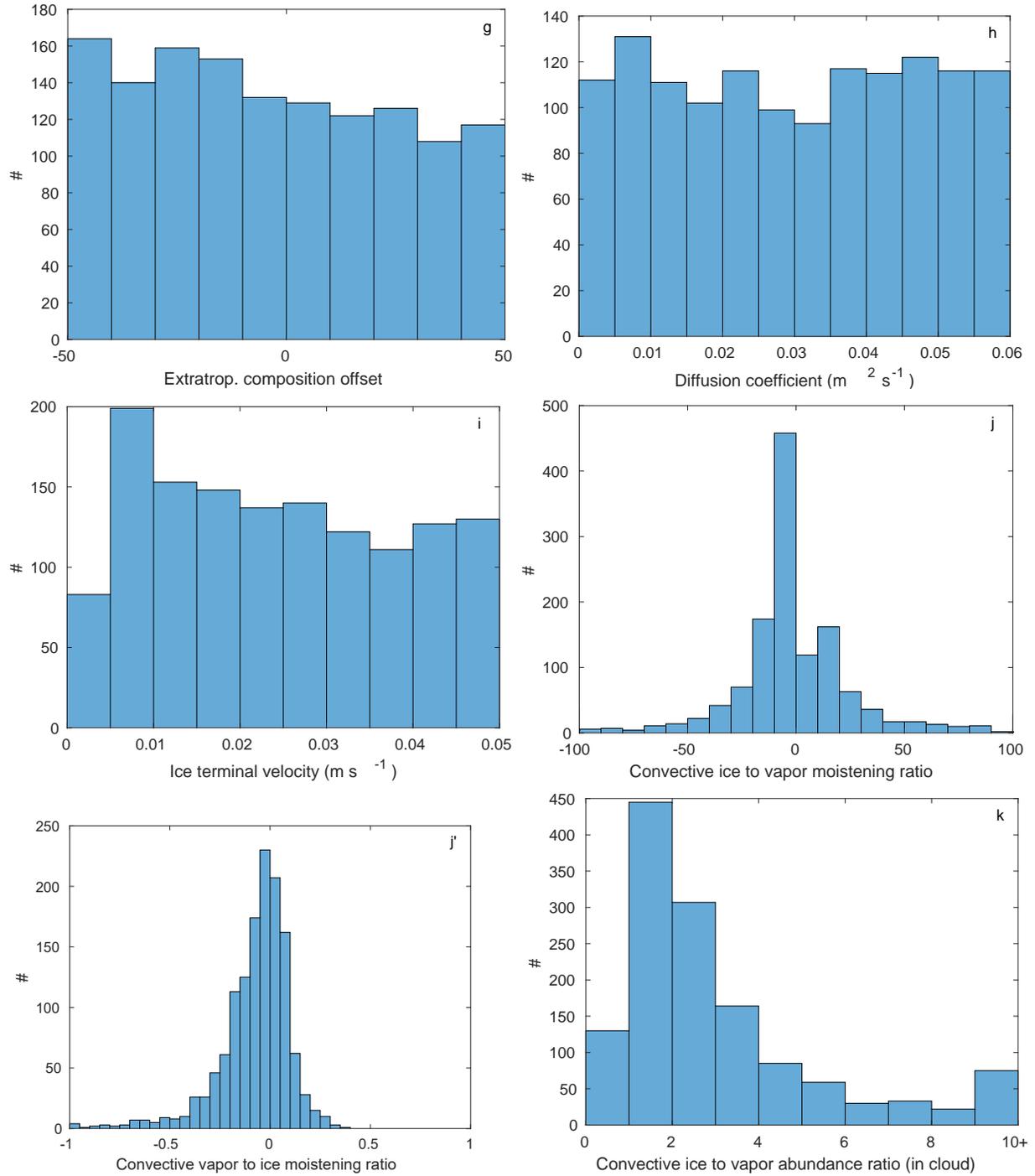

Figure S9: CONTINUED FROM S8. Histogram of (g): extratropical vapor isotopic composition offset; (h): value of the diffusion coefficient; (i): value of the ice terminal velocity; (j): ratio of ice to vapor contributions in convective moistening $(e/(D(r_{vc} - r_v)))$; (j'): same as (j) but reversed (vapor to ice contributions); (k): ratio of ice to vapor abundance at the top of convective towers $(e/(Dr_{vc}))$.

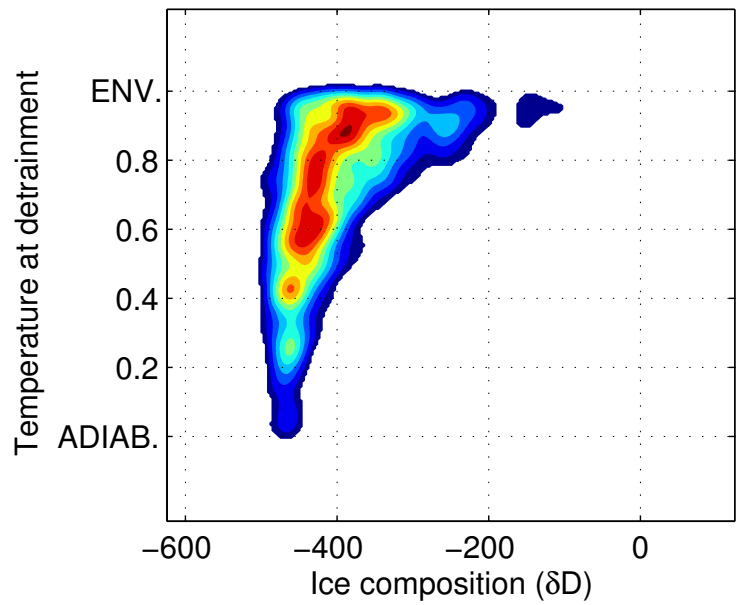

Figure S10: Joint probability density at 16.5 km of sublimating ice composition and temperature index at detrainment for those cases that yield ice within +/- 50% of CALIOP at this altitude.



\end{document}